# RF-PUF: Enhancing IoT Security through Authentication of Wireless Nodes using In-situ Machine Learning

Baibhab Chatterjee, *Student Member, IEEE*, Debayan Das, *Student Member, IEEE*,
Shovan Maity, *Student Member, IEEE* and Shreyas Sen, *Senior Member, IEEE*

*Abstract*— Traditional authentication in radio-frequency (RF) systems enable secure data communication within a network through techniques such as digital signatures and hash-based message authentication codes (HMAC), which suffer from key-recovery attacks. State-of-the-art IoT networks such as Nest also use Open Authentication (OAuth 2.0) protocols that are vulnerable to cross-site-recovery forgery (CSRF), which shows that these techniques may not prevent an adversary from copying or modeling the secret IDs or encryption keys using invasive, side channel, learning or software attacks. Physical unclonable functions (PUF), on the other hand, can exploit manufacturing process variations to uniquely identify silicon chips which makes a PUF-based system extremely robust and secure at low cost, as it is practically impossible to replicate the same silicon characteristics across dies. Taking inspiration from human communication, which utilizes inherent variations in the voice signatures to identify a certain speaker, we present RF- PUF: a deep neural network-based framework that allows real-time authentication of wireless nodes, using the effects of inherent process variation on RF properties of the wireless transmitters (Tx), detected through in-situ machine learning at the receiver (Rx) end. The proposed method utilizes the already-existing asymmetric RF communication framework and does not require any additional circuitry for PUF generation or feature extraction. The burden of device identification is completely shifted to the gateway Rx, similar to the operation of a human listener's brain. Simulation results involving the process variations in a standard 65 nm technology node, and features such as LO offset and I-Q imbalance detected with a neural network having 50 neurons in the hidden layer indicate that the framework can distinguish up to 4800 transmitters with an accuracy of 99.9% (≈ 99% for 10,000 transmitters) under varying channel conditions, and without the need for traditional preambles. The proposed scheme can be used as a stand-alone security feature, or as a part of traditional multi-factor authentication.

*Keywords*—Radio Frequency, Authentication, PUF, Machine Learning, Deep Neural Network, Internet-of-Things (IoT), Artificial Neural Networks (ANN), Security, Device Signatures

Manuscript received Feb 15, 2018; revised April 24, 2018; accepted June 5, 2018. Date of current version June 17, 2018. This work is recommended by Associate Editor: Dr. Mohamed Kheir and Editor in Chief: Dr. Sherman Shen.

This work is supported in part by National Science Foundation (NSF) SaTC, CNS Grant No. 1719235 and Semiconductor Research Corporation (SRC) Grant No. 2720.001.

The authors are affiliated with the School of Electrical and Computer Engineering (ECE), Purdue University, West Lafayette, IN 47907, USA. E-mail: {bchatte, das60, maity, shreyas}@purdue.edu.

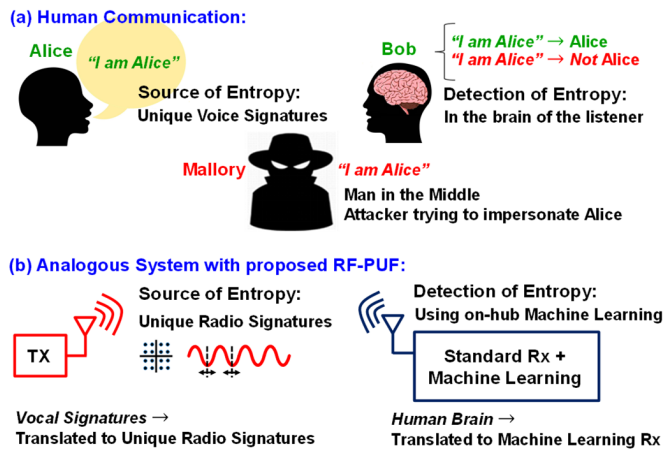

Fig. 1. (a) Authentication in Human Voice Communication: Bob (the receiver) can identify Alice (the transmitter) based on the unique voice signatures, and not based on the contents of what Alice speaks. Mallory (the impersonator) can also be identified (as not Alice), since his unique voice signatures would be different from Alice. (b) Analogous System that utilizes the proposed RF-PUF framework for secure radio communication.

## I. INTRODUCTION

### A. Background and Motivation

The advancements in sensor electronics, wearable technology and mobile computing/communication platforms have resulted in an unprecedented data deluge in the domain of Internet-of-Things (IoT). According to CISCO's Visual Networking Index (VNI) based Global Mobile Data Traffic Forecast, machine-to-machine (M2M) communication systems are expected to have about 27.1 billion connected devices by 2021 [1]. Due to their inherent mobile nature, these devices perennially operate under untrusted environmental conditions and are exposed to a number of potentially malicious attacks. The development of mobile hardware security has been comparatively slower than the improvements in computation power [2]. When these devices are required to be securely authenticated using a symmetric-key implementation, a secret key is usually placed in a non-volatile memory (NVM) or a battery-backed SRAM and is subsequently used in a digital signature or hash-based encryption. However, these techniques are vulnerable to key-hacking (through invasive/semi-invasive/software/side channel attacks) and come with significant area and power overhead for the NVM/SRAM implementation. The widely used OAuth 2.0 protocol [3] for current IoT devices suffer from cross-site-recovery-forgery

(CSRF) attacks, and may eventually become cumbersome as the number of devices per user grows (OAuth requires the user to manually authenticate every device in the network). Because of these reasons, Physical Unclonable Functions (PUF) have emerged as a promising alternative/augmentation which exploit manufacturing process variations to generate a unique and device-specific identity for a physical system [4]-[7]. PUF implementations are simpler than memory-based solutions as they consume significantly less energy and chip area than expensive cryptographic hardware such as secure hash algorithm (SHA) or public/private key encryption with NVM/SRAM that may also require anti-tamper mechanisms to detect invasive attacks.

PUFs are usually classified into strong and weak PUFs depending on the number of challenge-response pairs (CRP) that they can handle. Weak PUFs support a small number of CRPs which are linearly related to the number of components used to build the PUF. Strong PUFs, on the other hand, support a large number of CRPs such that polynomial time attacks become infeasible. These type of PUFs are usually employed in device authentication applications [2].

IoT systems can significantly benefit from a PUF-based authentication protocol, wherein the physical characteristics of each transmitter in the wireless sensor network can be analyzed and stored in a secure server as a general technique, thereby augmenting or replacing traditional key-based authentication schemes. In modern Digital Communication, ideal digitally modulated data pass through device-dependent unique analog/RF impairments (for example, frequency error/offset and I-Q imbalance) in the transmitter chain, which are compensated for at the receiver. These process-dependent non-idealities are already present in the wireless communication signal path, and are traditionally discarded/minimized as unwanted non-idealities. In RF-PUF, we embrace those existing non-idealities through an in-situ light-weight machine learning engine at the receiver side, that extracts the 'entropy' and creates a 'strong PUF' to securely identify the transmitters. This is similar to the inherent authentication in human voice communication as shown in Fig. 1(a). Bob (the receiver) can identify Alice (the transmitter) based on the unique voice signatures, and not based on the contents of what Alice says. Mallory (the impersonator) can also be identified as his unique voice signatures would be different from Alice. The source of entropy is in the vocal signatures of speaker (no extra hardware for entropy extraction), whereas the decoding of the entropy is in the listener's brain (heavy lifting at the receiver). Fig. 1(b) shows the analogous system (RF-PUF), wherein the unique signatures in each transmitter are used for device identification in the brain of the receiver, represented by the machine learning hardware. The entire system is envisioned in Fig. 2 in presence of channel impairments, with each of the transmitters inherently working as a PUF instance [8]. Such an implementation not only helps in authentication of physical nodes in resource-constrained IoT environment, but also enables applications such as intrusion detection, forensic data collection, defect detection/monitoring and body-connected biosensors as shown in [9]-[12].

*B. Our Contribution*

We have previously demonstrated the methods of process-detection in wireless radios [13]-[15] and the effect of variations in analog/mixed-signal/RF properties of such radios to adapt it for zero-margin operations [16]-[18]. In this paper, we build on that expertise to identify radio instances based on their inherent signatures automatically imparted on communicated signal, leading to a detailed analysis of the PUF properties of radios for enhanced physical layer security. The major *contributions* of this work are as follows:

1) A conceptual development of RF-PUF is presented for an *asymmetric* IoT network, which consists multiple low-cost, low-power, distributed transmitters, and a single central hub as a receiver. To the best of the authors' knowledge, this is the first work in this area for low-cost, preamble-less, intrinsic PUF-based authentication of IoT nodes. This is in contrast to traditional RF fingerprinting methods which are usually preamble-based and/or software defined as discussed in the next section.

2) Enabling RF-PUF operation without any extra-hardware at the resource-constrained IoT node. As described in Section III, RF-PUF does not require any additional on-chip/off-chip circuitry for PUF implementation at the Tx. The proposed scheme makes use of inherent variations resulting from factors such as process variability (on-chip) and component tolerance (on-board) for each transmitter. A method to compensate for non-ideal Rx signatures is also proposed in Section V.

3) Conforming to our earlier work on on-hub analytics where we proposed a method of staged inference using conditional deep learning [12], a light-weight machine learning framework is developed in the current work, which compensates for receiver non-idealities, and accounts for both data variability and channel variability at the same time. Since this is a non-linear multidimensional classification problem, an Artificial Neural Network (ANN) is employed as a learning engine. Simulation results with $\approx 10000$ transmitters demonstrate $\approx 99\%$ accuracy using supervised learning which proves the practical feasibility of RF-PUF for IoT-based applications targeted towards small to medium-scale smart systems with about a thousand devices connected to a single gateway receiver.

In essence, the proposed method lends the biggest benefits in asymmetric smart networks as 1) no extra hardware is required at the resource-constrained IoT nodes, while the heavy-lifting is performed by the gateway receiver, which is similar to a listener's brain, 2) the method can be employed as a stand-alone physical-layer security feature, or for multi-factor authentication, in conjunction with network-layer, transport-layer and application-layer security features.

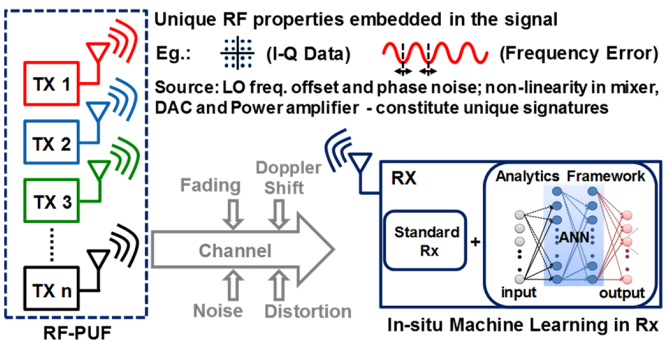

Fig. 2. Visualization of RF-PUF at a system level [8]

The remainder of the paper is organized as follows: Section II lists the most recent developments in the area of PUF and RF authentication, along with their applications in the relevant domains. Section III presents the architecture of the proposed PUF in detail, while Section IV illustrates the performance metrics of the proposed method. The simulation results, along with the security aspects are analyzed in detail in Section V. Finally, Section VI summarizes this paper and points to the future directions.

## II. RELATED WORK

The notion of Silicon PUFs was introduced by Gassend et al. in [19], where the authors illustrated the use of PUF in anti-counterfeiting applications by measuring the intrinsic delays in a self-oscillating circuit. However, the additional requirement of robustness in a large sample pool led to the inclusion of various error-correction mechanisms at the system-level [20]-[23]. The error correction improves the system reliability at the cost of additional software/hardware burden on the PUF implementation. As shown in [2], many of these techniques tend to leak the secret keys that are used to generate the syndrome bits. In such a scenario, a higher number of PUF bits are generated first, and then are down-mixed to increase entropy.

RF fingerprinting has been a popular method to automatically identify the wireless nodes in a network by using the time and frequency-domain properties extracted during transmitter power-on [24]-[27]. While the transient properties are consistent, they offer acceptable classification accuracy only when the beginning and the end of the transient can be reliably identified [10]. Moreover, the analysis of transient properties require very high oversampling rates (500 MS/s in [25] and 50 GS/s in [26]) which pose significant power and precision requirements that lead to expensive receiver architectures. An alternative and less adopted approach involves the use of steady-state properties of the transmitters, which are extracted after the communication loop transients are settled. However, the steady-state signal is data-dependent in different transmissions (with different bit-streams) which makes it unsuitable for identification purpose. For this reason, previously reported literature [9]-[10] use fixed random channel access (RACH) preambles along with techniques such as spectral averaging or matched filtering to correctly identify the transmitters. The steady-state analysis method is relatively unexplored, but is promising as it does not require sophisticated and power-hungry receiver architectures. More importantly, the steady-state portion of the signal can easily be identified as opposed to the transient states during power-on.

In this study, we combine the concept of PUF with RF fingerprinting to develop a system architecture that utilizes RF properties of the transmitters to identify nodes using an in-situ machine learning framework at the receiver. A preamble-less steady state approach is adopted which is implemented by training the learning sub-system with multiple data streams and with different channel conditions. Although the proposed approach utilizes manufacturing variabilities in the transmitters for device identification (similar to state-of-the-art RF fingerprinting [28]-[31]), *RF-PUF is unique from RF fingerprinting in 4 different aspects*: (1) the operation for RF-PUF is not preamble-based, (2) unlike transient mode RF fingerprinting, RF-PUF does not require high oversampling ratio at the receiver, (3) RF-PUF utilizes significantly higher dimensionalities in the feature space than steady-state RF fingerprinting that gives rise to its strong PUF properties (Section III C.(f)) while providing justification for the nomenclature, (4) RF-PUF compensates for the non-ideal receiver signatures (Section V. A), thus allowing a large number of devices/challenge-response-pairs. Machine learning had been used in prior work [29]-[30] for device identification, but the proposed work also identifies machine learning as a solution to the practical challenge of receiver signature compensation that often limits larger system implementations. As compared to previous implementations such as the RF-DNA [32], RF-PUF does not require any additional analog/RF hardware for PUF implementation at the Tx as the features are selected such that feature generation and extraction is ingrained in the transceiver operation (RF-DNA involved measuring the reflected/refracted EM waves based on the 3D-positioning of scattering antennas which are different for each RF unit). Moreover, error-correction and noise cancellation measures are also intrinsic to the transceiver architecture, which increases the reliability of the proposed RF-PUF without the need for any specific error correction mechanism dedicated for the PUF operation.

## III. PROPOSED PUF

This work primarily focuses on a technique of authenticating devices within a low-cost swarm of IoT nodes, which can have significant variation from node to node. If all the components are tightly controlled during fabrication and manufacturing, the standard deviation of variation will be less and the number of unique devices which are correctly identified will reduce. Interestingly, that will increase the cost of fabrication significantly and hence it is cheaper and easier to embrace the non-idealities (up to a point where it does not affect the overall performance) which justifies the use of RF-PUF.

### A. Features utilized in RF-PUF implementation

As described below, the PUF properties of the system originate from the manufacturing variability of the Tx(s). The identification of each node is performed in the Rx sub-system which extracts multiple features from the received signals.

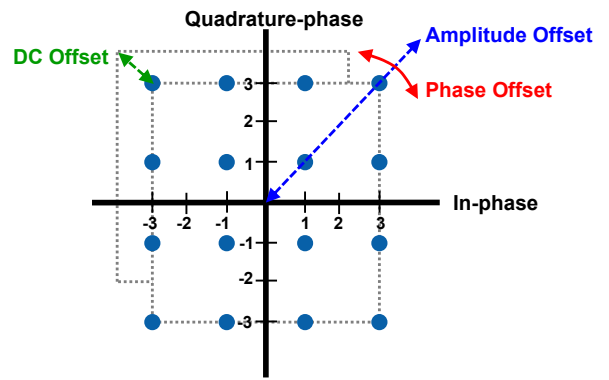

Fig. 3. DC, Amplitude and Phase imbalance in 16-QAM

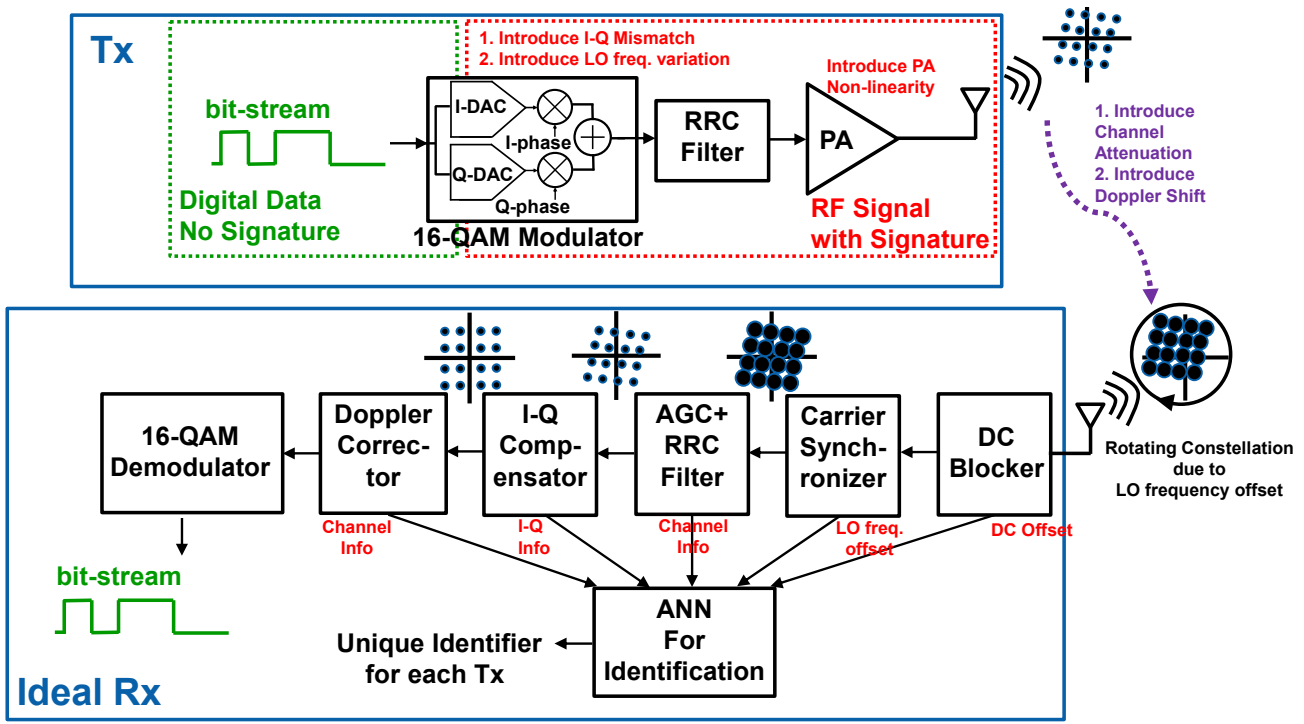

Fig. 4. System-level simulation setup involving Transmitter and Receiver for RF-PUF implementation

*(a) Frequency features:* Every transmitter has its unique frequency offset with respect to the ideal carrier frequency because of inherent variations in the local oscillator (LO). This offset has been used as a prime feature for device identification in [10] and [33]. The allowable limits for the frequency offsets can be different for different established standards. For example, the frequency error must be within $\pm 25$ ppm of the center frequency for the IEEE 802.11b standard for WiFi (and within $\pm 40$ ppm for IEEE 802.15.4 which is one of the preferred standard for IoT). This corresponds to a normal distribution with standard deviation ($\sigma$) of 20.1 kHz on either side of the 2.412 GHz center frequency for 802.11b. With a high-quality reference clock (low jitter with zero mean) in the Rx, frequency offsets for multiple Tx(s) in the system can be calculated. In our implementation, a carrier synchronizer module (which already exists in a standard receiver for LO offset compensation) is employed in the receiver that finds the frequency offset and compensates for it. The ppm value of the offset is provided to a 3-layer machine-learning framework for identifying the device.

*(b) I-Q features:* The amplitude and phase mismatch between the in-phase (I) and quadrature (Q) components of the transmitted signal is unique for different transmitters. Fig. 3 shows the nature of these imbalances and how they can impact the constellation diagram at the receiver for a 16-QAM (quadrature amplitude modulated) signal. Other transmitter variabilities such as power-amplifier (PA) back-off and gain variations also affect the constellation. Compressive nonlinearity affects the outer symbols in the constellation more than the inner symbols. Hence it is necessary to extract amplitude and phase information for all symbols in the constellation. These features, along with the frequency errors from each transmitter has the potential to uniquely identify each transmitter in the network.

*(c) Channel features for compensation:* The communication channel introduces time and frequency dependent variations in various forms such as attenuation, distortion and Doppler shift. To establish reliable operation of the RF-PUF, these channel properties need to be estimated and compensated. For this purpose, an automatic gain control (AGC) block, a Root-Raised Cosine (RRC) filter and a Doppler corrector is employed as indicated in Fig. 4. The RRC filter, along with the AGC module, helps reducing inter-symbol interference (ISI) and provides a measure of the channel attenuation to an ANN. Similarly, the Doppler corrector module estimates and corrects the amount of Doppler shift due to any physical movement of the transmitters and receivers in the network, and provides the information to the ANN for channel compensation.

*B. Communication System Example in RF-PUF: 16-QAM*

Fig. 4 shows the entire transceiver system for RF-PUF authentication. The 16-QAM transmitter does not have any additional circuitry for PUF implementation. The receiver, on the other hand, has multiple stages for RF signal processing and simultaneously performs feature extraction at various stages. A simple 3-layer neural network takes the extracted features as inputs and identifies the transmitters based on training data.

*C. Properties of the RF-PUF*

The proposed system has the necessary and sufficient properties of a PUF [5] that makes it suitable for security/authentication applications:

(a) *Constructability*: A PUF class $\mathbb{P}$ is constructible if a random PUF instance ($p_r \in \mathbb{P}$) can be created by invoking a particular creation procedure, $\mathbb{P}.\text{Create}: p_r \leftarrow \mathbb{P}.\text{Create}$

RF-PUF aims to exploit the technical limitations that exist in the physical process of fabricating the RF transmitters. Hence, the manufacturing process itself serves as the creation procedure $\mathbb{P}$. Create for each of the PUF instances (RF transmitters).

(b) *Evaluability*: A constructible PUF class $\mathbb{P}$ is evaluable if for a random PUF instance ($p_r \in \mathbb{P}$) and a random challenge ($x$), it is possible to evaluate a response $y$: $y \leftarrow \mathbb{P}.\text{Eval}[p_r(x)]$

For RF-PUF, $x$ is the challenge input bit-stream to the transmitter, and $y$ is the unique analog response for each PUF instance. The uniqueness in the response can be attributed to multiple features, due to the die-to-die and within-die variations.

(c) *Reproducibility*: A PUF class $\mathbb{P}$ is reproducible/reliable if it is evaluable and if the probability of intra-PUF variation being lower than a system-defined small number is very high.

For the RF-PUF, the measure of reproducibility can be defined as the difference ($D_{\text{intra}}$) between two distinct evaluations ($y(x), y'(x)$) of a particular PUF on the challenge $x$.

$$D_{\text{intra}}(x) \cong dist[y_1(x), y_1'(x)]$$

$D_{\text{intra}}$ is the intra-chip (intra-PUF) distance that serves as a metric to measure the resilience of the RF-PUF to varying environmental conditions. Reproducibility/reliability is also a measure of stability as $D_{\text{intra}} = 0\%$ in an ideal scenario.

(d) *Uniqueness*: A PUF class $\mathbb{P}$ is unique if it is evaluable and if the probability of inter-PUF variation being higher than a system-defined large number is very high.

For the RF-PUF, the measure of uniqueness can be defined as the difference of the responses between two PUF instances ($y_1(x), y_2(x)$) evaluated with the same challenge $x$.

$$D_{\text{inter}}(x) \cong dist[y_1(x), y_2(x)]$$

(e) *Identifiability*: A PUF class $\mathbb{P}$ is easily identifiable if it is reproducible as well as unique, and if the probability of intra-PUF variation being lower than inter-PUF variation is very high.

$$\text{Prob}(D_{\text{intra}} < D_{\text{inter}}) \approx 1$$

As will be seen in Section IV, RF-PUF simultaneously exhibits reproducibility, uniqueness and identifiability. In the simulation results shown in Section IV, the worst case $D_{\text{inter}}$ (3.9 ppm : geometric mean of ppm variations over all features) was found to be larger than the corresponding $D_{\text{intra}}$ (2.9 ppm : geometric mean of ppm variations over all features) for 1000 transmitters. These properties coupled with the physical unclonability and unpredictability of the silicon manufacturing process makes the RF-PUF implementation practically feasible.

(f) PUF Strength: The challenge for RF-PUF is a digital data sequence, while the response contains the analog features embedded in actually transmitted RF signal (and hence in the received signal) which is unique for each Tx. Since the challenge-response-pairs (CRPs) consist features which are real-valued analog numbers, the total number of CRPs for $m$ distinct analog/RF features will be $\mathbb{R}^m$ where $\mathbb{R}$ represents all values in the real number space within a range of $\pm 3\sigma$ around the mean. When each of these features is quantized using a 16 bit ADC, for example, $\mathbb{R}^m$ translates to $2^{16m}$ which is a large number even for small values of $m$ (3-10). Hence, if a probabilistic polynomial time (PPT) adversary knows the responses from $k$ transmitters, it is possible to predict the response for the $(k + 1)$-th transmitter only with a negligible probability of $\left(\frac{1}{2^{16m}}\right)$. This property makes RF-PUF a '*strong PUF*' which is suitable for authentication applications [2].

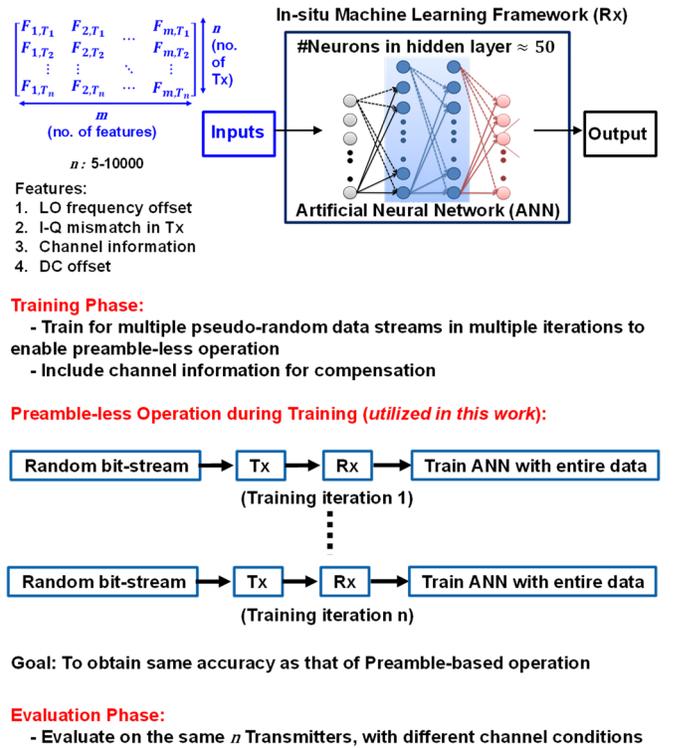

Fig. 5. Preamble-less Training for the ANN. This method enables on-the-fly authentication without knowing the expected bit-stream from the Tx [8].

### D. Training the ANN for Device Identification

The 3-layer ANN is trained in multiple iterations with different pseudo-random bit-streams (PRBS) to account for data variability in evaluation stage and to *enable preamble-less operation*. The hyper-parameters, including the number of training iterations (shown in Fig. 6(c), that represent sampling and collection of data streams under dynamic channel conditions which vary slowly during a single evaluation period but can change significantly from one evaluation to another), were optimized aggressively through scaled conjugate gradient backpropagation algorithm, using variable number of epochs and a target training-set-error. For the data presented in subsequent sections, the number of training iterations was set to 10, with a stream length of 30,000 bits. The training dataset was presented to the ANN as a $n \times m$ feature matrix as shown in Fig. 5. The extracted channel conditions for the 10 iterations were also presented as features so that the network learns to compensate for the channel variations. The number of neurons in the hidden layer was varied and was optimized for performance (detection accuracy). During Evaluation phase, a different pseudo-random data-stream from any of the $n$ transmitters is provided to the ANN in presence of a different channel condition. This mode of training helps the neural network to learn both data variabilities and channel variabilities, thereby leading to a robust system.

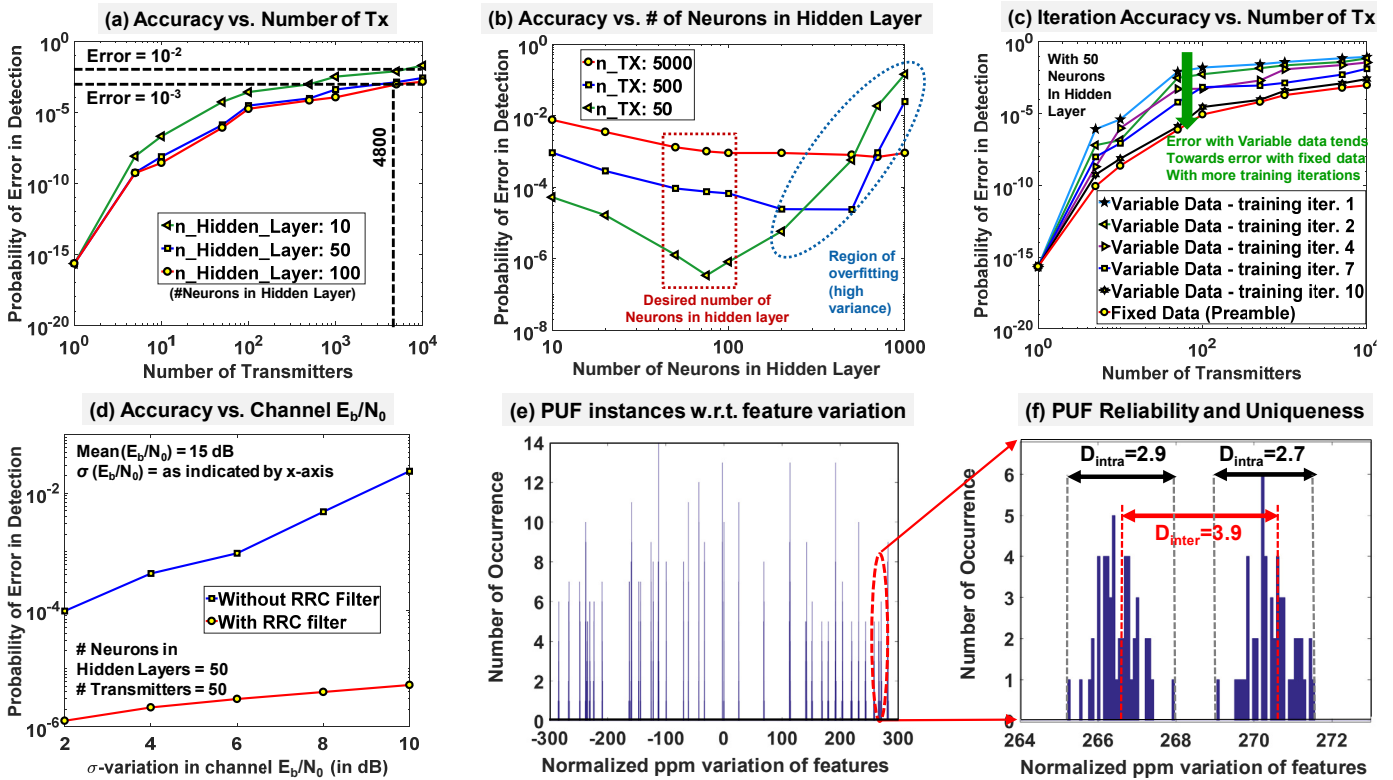

Fig. 6. (a) Probability of False detection as a function of the total number of Tx in the system. (b) Probability of False detection as a function of the number of neurons in the hidden layer of 3-layer ANN. (c) Reduction in prediction error with increased number of training iterations. Each iteration trains the network with a different bit-stream and different channel conditions. (d) Probability of False detection as a function of the standard deviation ($\sigma$) of the $E_b/N_o$ in the channel in presence of additive white gaussian noise (AWGN). (e) PUF instances with 1000 Tx (50 instances for each Tx, with different channel conditions) and 50 neurons in hidden layer w.r.t. geometric mean of normalized ppm variation of features. (f) Worst case reliability ($D_{intra}$) and uniqueness ($D_{inter}$) for the 1000 transmitters.

## IV. PERFORMANCE METRICS

The proposed PUF is simulated using the Neural Network toolbox in MATLAB, with a 16-QAM modulation scheme. The manufacturing process variations of a standard 65 nm technology is included during the simulations to model the statistical variability in the range of ($\mu \pm 3\sigma$). A total of 10,000 PUF devices were simulated under varying channel conditions. Table I shows the mean and standard deviation of the transmitter features and channel features that are used during simulations. The LO frequency and frequency error follows the IEEE 802.11b standard. I-Q imbalance has a mean value of 0, while the linearity of the power amplifier (PA) is defined by the back-off value (w.r.t. 1 dB compression point). $E_b/N_0$ is the ratio of energy per bit and noise which defines the signal to noise ratio (SNR) at the receiver.

TABLE I
TRANSMITTER AND CHANNEL FEATURES USED FOR SIMULATION [34][35]

| Feature | Average ($\mu$) | Standard Deviation ($\sigma$) |
|---|---|---|
| LO Frequency | 2.412 GHz | 20.1 kHz (8.3 ppm) |
| I-Q Amplitude Imbalance | 0 dB | 1 dB |
| I-Q Phase Imbalance | 0° | 5° |
| PA back-off (linearity) | 30 dB | 1 dB |
| $E_b/N_0$ | 15 dB | 2 dB |
| Doppler Shift | 0 Hz | 1 Hz |

### A. Probability of False Detection

Fig. 6(a) shows the probability of false detection of a Tx as a function of the total number of transmitters (n_Tx) in the network, which indicates that the error is < $10^{-3}$ up to 4800 transmitters, and < $10^{-2}$ for 10,000 transmitters. Fig. 6(b) illustrates the probability of false detection of a transmitter as a function of the number of neurons in the hidden layer of the ANN (n_Hidden_Layer). It is to be noted from both Fig. 6(a) and Fig. 6(b) that the probability of an error in detection does not reduce much when the n_Hidden_Layer is more than 50. Increasing the size of the neural network beyond this limit will cause overfitting and will increase the power and area cost without significant performance benefit.

Fig. 6 (c) shows the effect of training in multiple iterations with variable data as compared to training with a fixed data-stream (preamble). In case of preamble-based training, the ANN is trained using only the fixed headers in the data-stream. In our implementation, conversely, the ANN is trained using variable data-stream in multiple iterations. As the number of training iterations increase, the performance of the system tends towards the performance achieved in fixed-preamble case. This overhead of additional training iterations are also useful in terms of learning the channel conditions and variabilities, as each iteration has a different channel condition which the network learns to compensate.

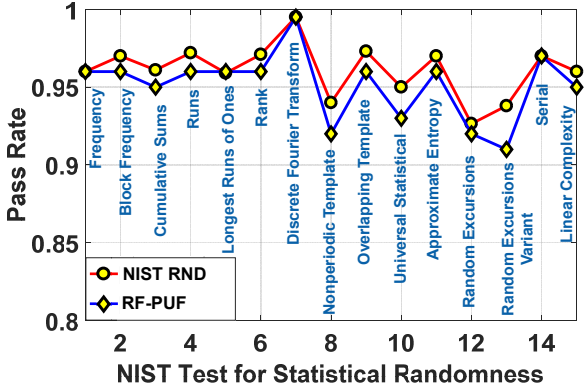

Fig. 7. Results of randomness test for RF-PUF using NIST Test Suite [38]

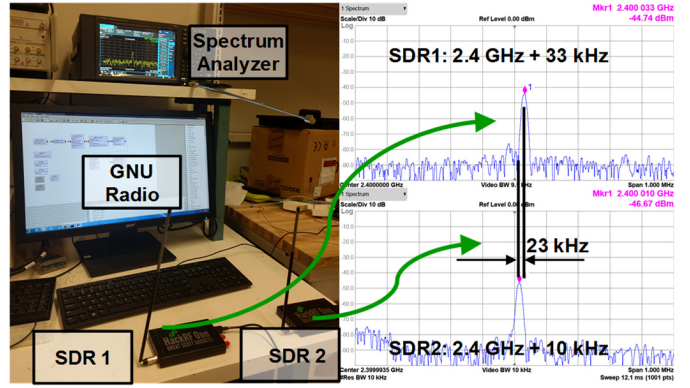

Fig. 8. Experimental Setup to extract physical transmitter properties. Two software-defined-radios (SDR) @ 2.4 GHz are shown in this setup, exhibiting a frequency difference of 23 kHz as non-ideality. The amplitude and phase imbalance is captured in the GNU Radio software after reception.

### B. Robustness to Noise, Dynamic Channel Variation and ISI

Even for short-range (< 30 m) communication, the channel is affected by noise/attenuation in the communication medium, interference, Doppler shift and fading, out of which the contribution of attenuation is the most dominant [36]. Fig. 6(d) illustrates the effect of channel attenuation on the probability of false detection. Without the RRC filter, inter-symbol interference (ISI) increases which leads to an error probability of $\approx 0.02$ in device identification when the standard deviation of channel $E_b/N_0$ is 10 dB (with mean $E_b/N_0 = 15$ dB). In presence of the RRC filter at the Rx, it is easier to extract features from the I-Q data for different transmitters, because of a reduction in ISI. As a result, probability of error reduces to as low as $10^{-5}$ for the range of variations shown in Fig. 6(d).

### C. Intra-PUF Hamming Distance (Reliability) and Inter-PUF Hamming Distance (Uniqueness)

Fig. 6(e)-(f) illustrates the Reliability and Uniqueness of RF-PUF w.r.t. the input features. It is to be noted that unlike a traditional PUF circuit that produces digital output, RF-PUF embeds the unique signature of the transmitter in the RF properties of the message signal, and hence the Intra-PUF and Inter-PUF distances are plotted using the normalized parts-per-million (ppm) variation of the features. To represent the ppm variations of multiple features on a single axis, a transformation of the feature spaces is required. Since the total number of possible PUF instances is proportional to $\prod_{i=1}^{m} N_i$ ($m$ = number of features, $N_i$ = number of possible values for feature $i$), it is intuitive to utilize the geometric mean of the ppm values of all the features for representing a PUF instance.

The worst case inter-PUF variation with 1000 transmitters is found from the simulations and is shown Fig. 6(f). The worst-case inter-PUF difference thus defined is very close to the intra-PUF difference which explains the high (in the range of $10^{-3}$ or higher) probability of false detection when number of transmitters are more than a few thousand (Fig. 6(a)).

### D. Randomness and Bias

For the guessing entropy to be low, the randomness of the PUF needs to be high, whereas the bias needs to be low. For the simulations shown in this paper, the *NIST recommended random bits* [37]-[38] are used to generate the features within a range of ($\mu \pm 3\sigma$), as shown in Table I. The resulting pass rates for the RF-PUF output (normalized geometric mean of feature values) is shown in Fig. 7, alongside the pass rates for NIST recommended random bits (NIST RND). Evidently, the pass rates for the PUF output are higher than 0.9 for each of the 15 tests, which matches closely with the NIST RND. Any bias in the PUF output is refuted by the Frequency test, which exhibits a pass rate > 0.95.

### E. Experimental Validation of RF-PUF: Physical Implementation with SDRs

To prove the feasibility of RF-PUF in hardware, an experimental setup (as shown in Fig. 8) is developed using software-defined-radios (SDR). The SDRs are ideally programmed to operate at 2.4 GHz, with zero I-Q imbalance. However, there will be both frequency error and I-Q non-idealities in the practical scenario. The frequency error is captured using a spectrum analyzer, while the I-Q imbalance can be captured by configuring one of the SDRs as a receiver, using the GNU Radio platform. Fig. 8 shows 2 SDRs, with a 23 kHz difference in their carrier frequencies. This difference in frequency can be detected at the receiver side from the down-converted I-Q data, by sensing the baseband signal over a duration which is inversely proportional to the difference in frequency. To generate enough number of unique classes that help validating the learning engine, multiple unique transmitters are artificially emulated from these SDRs by modifying the ambient temperature in a closed environment. Changing the temperature in discrete steps of 5℃ in the range of 0℃ -25℃ modifies the Tx properties, and the receiver considers every 5℃ change in the temperature as a different transmitter when it does not have any information about the temperature. The non-ideality information from the 2 SDRs, along with data from 8 other emulated transmitters are provided as input to the neural network framework (Section III.D), which detects all 10 transmitters correctly. This confirms that the neural network can identify multiple transmitters with a small difference in their RF features. In a more realistic scenario, the features from the transmitters need to be automatically calibrated/compensated at the receiver w.r.t. different temperature and supply voltages before the classification using the ANN. For that, the system would require a temperature and supply sensing mechanism, and additional pre-processing at the on-board application processor. Alternatively, the temperature and voltage information can be directly provided to the neural network so that it learns to compensate for the variations, which would be included as a future work.

## V. Discussion on Rx Signatures, Attack Models, Security and Robustness

### A. Compensating Rx Signatures

It is evident from Fig. 6(a) that the machine learning framework can support up to 10,000 transmitters with a probability of false detection $< 10^{-2}$ and up to 4800 transmitters with a probability of false detection $< 10^{-3}$. Fig. 6(b) indicates that the required number of neurons in the hidden layer should be within 50-100 to successfully detect the transmitters, which can be implemented on a processor in case of a software-defined-radio environment. While this is promising in terms of the conceptual feasibility, the receiver non-idealities will pose significant implementation constraints on the system.

For the initial system simulation, an ideal receiver has been assumed which does not insert any signature of its own into the feature set extracted from the transmitters. However, in a practical scenario, the receiver will alter the signatures in the received signal. This can be taken care of using a loopback analysis to find the receiver properties [39], post which the system can be incrementally trained for such receiver non-idealities. Alternatively, the receiver sub-system can be corrected pre-emptively for non-idealities (values of which are obtained through the loopback test) using a second set of LO offset compensator (carrier synchronizer), AGC and I-Q compensator modules as shown in Fig. 4.

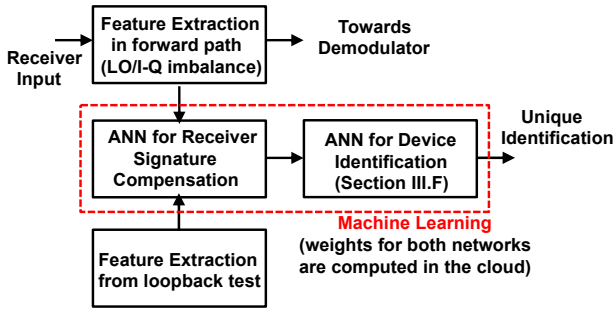

Fig. 9. One possible method that accounts for receiver non-idealities while performing device identification. A neural network performs the receiver signature compensation before device identification.

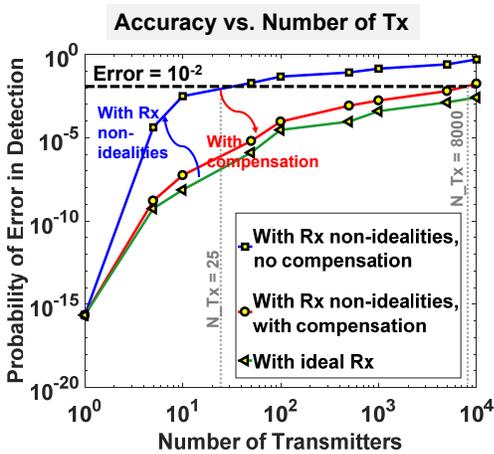

Fig. 10. Probability of False detection as a function of the total number of transmitters in the system, with and without receiver signature compensation.

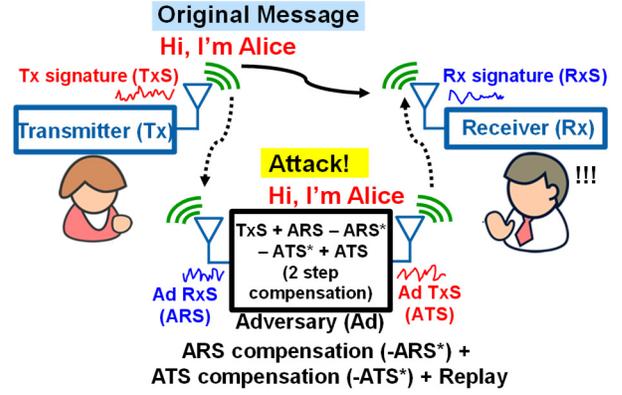

Fig. 11. One possible Replay Attack model: Adversary needs to perform its own receiver signature compensation (-ARS*) and transmitter signature compensation (-ATS*). The necessary conditions to successfully replay the message from Alice with the original transmitter signature (TxS) are ARS=ARS* and ATS=ATS*.

Fig. 9 shows the training-based method for compensating the receiver signatures. A second neural network learns the functions to be employed on individual features to compensate the non-idealities, and then performs the linear transformations on each of those features. The compensated feature vector is then provided to the original ANN that performs the device identification. Fig. 10 illustrates the effect of this compensation method. When the number of transmitters is more than 25, the probability of false detection increases from $10^{-7}$ in the ideal receiver case to $10^{-2}$ in the practical receiver case without compensation. With compensation, this value again reduces to $10^{-6}$. When the number of transmitters is > 8000, the error is about $10^{-2}$ in the case with receiver compensation, which is sufficient for most smart-home applications.

### B. Possible Attack Models

In [40], the possible attack mechanisms on a strong PUF are classified into two primary categories: (1) the PUF re-use model and (2) the malicious PUF model. In most practical applications, the attack usually comprises of a combination of the two models. The PUF re-use model is based on the scenario that an adversary can have repeated temporary physical access to the PUF when the PUF is communicating with an authenticating medium. This presents the adversary an opportunity to model and replay the responses. On the other hand, the malicious PUF model assumes that either the PUF responses can be simulated using a software algorithm, or the adversary can have direct access to all the CRPs through a built-in logger program/implanted Trojan. Since RF-PUF does not store any digitally encoded signature, it does not suffer from the malicious PUF model. However, it can potentially suffer from PUF re-use models, as described in the next part of the discussion.

(a) Replay attack model:

Replay attacks are theoretically possible in the proposed RF-PUF based authentication system, as a wireless transmitter can send identical digital data-streams repeatedly (in other words, PUF re-use cannot be avoided). This scenario is presented in Fig. 11. Alice, the Tx wants to send a message, "Hi, I'm Alice" to the gateway receiver. This message inherently contains the Tx signature (TxS). We assume that the adversary (Ad) intercepts the transmitted data and in the process, adds the attacker's receiver signature (ARS) to the signal. Subsequently, Ad compensates for the ARS using – ARS*, which require high-speed and high resolution circuits that can negate minuscule changes in voltage and time. To successfully mimic Alice's signatures, Ad needs to compensate for its own Tx signatures (ATS) as well. This also utilizes high speed and high resolution circuits and compensates ATS using –ATS*. If ARS=ARS* and ATS=ATS*, the adversary can mimic Alice by sending the same message and same signatures. However, implementing both ATS and ARS compensation would require expensive ADCs and DACs for achieving negligible residual errors.

One may argue that the machine-learning-based gateway receiver compensation shown in Section V A. could also need similar high-resolution ADCs. However, imperfect compensations in the gateway receiver results in a small residual error, which only introduces a constant shift in the detection threshold for all the transmitters. On the other hand, for the attacker model, both ARS and ATS compensations have to be accurate. Otherwise, the replay attack is imperfect as the replayed signal does not mimic Alice's signature, and hence the adversary may not be successful in impersonating Alice. This limitation makes the replay attack extremely costly in the RF-PUF scenario.

(b) Machine Learning (ML)-based modeling attacks:

The proposed system can also suffer from a machine-learning attack as an external attacker can model the responses (using a separate learning engine) from RF-PUF through repeated access to the transmitted data.

However, it must be noted that the proposed system utilizes a supervised learning algorithm to uniquely identify the PUF devices. For an external attacker, mimicking a particular transmitter will translate to an unsupervised learning (clustering) problem, and hence the modeling accuracy will directly depend on the ratio of the number of CRPs accessed by the attacker to the total number of CRPs [41]. The modeling time, on the other hand, will depend on the number of CRPs that the adversary has access to. Hence, there exists a trade-off between the modeling time and accuracy, which is described with an example in the next section.

C. *Countermeasures against PUF re-use Attacks*

In [42], the authors implemented a strict one-time-usability protocol to avoid PUF re-use and modeling attacks. However, for our application, one-time-use is not feasible, as the transmitters (i.e. the PUF instances) need to send data whenever they are required to. As shown in [40], an alternative way to thwart such attacks is to simultaneously incorporate two additional properties during PUF implementation: (1) erasability and (2) certifiability. Erasability requires that the single responses from the PUF would be made impossible to read back without affecting other responses. This facilitates tamper-detection, but requires logical reconfigurability and additional control circuitry to be implemented on-chip. Certifiability, on the other hand, signifies that a PUF response must be checked offline (without any external "Trusted Authority") for possibilities of tampering, or any change in the PUF properties. This also requires additional circuitry in each PUF instance. A detailed analysis of the additional circuitry required for erasability and certifiability is out of scope of this work.

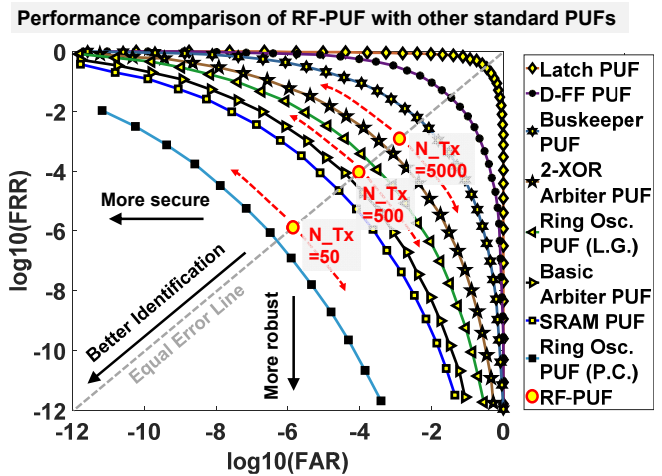

Fig. 12. Comparison of RF-PUF with various other PUFs available in literature. The values for False acceptance rate (FAR) and False Rejection Rate (FRR) for other PUFS are taken from [44].

In the absence of additional circuits for erasability and certifiability, RF-PUF can suffer from external ML attacks. In this section, we perform an estimate of the training time for an ML attack in order to demonstrate its practical limits for a problem with high dimensionality. As shown in Section III C.(f), the number of CRPs for the RF-PUF is $2^{16m}$ where $m$ is the number of features considered. For a nominal value of $m=5$, the the number of CRPs become $2^{80} \approx 10^{24}$. [43] shows that a machine-learning based attack model on $10^6$ CRPs takes 267 days to get completed on an INTEL Quadcore Q9300 processor. $10^{24}$ CRPs will result in significantly higher training time, leading to a wait time of several years before completing the attack.

D. *Security and Robustness*

In Fig. 12, RF-PUF is compared with various other PUFs in state-of-the-art literature [44]. Two well-known metrics – false acceptance rate (FAR) and false rejection rate (FRR) are used to define the security and robustness, respectively, for the PUFs. A low value for both FAR and FRR ensures that the authentication is both secure and robust. However, there is a trade-off between these two quantities which can be controlled by the detection threshold of PUF inferencing mechanism (the machine learning framework for RF-PUF). If the inferencing is done by a threshold which is much higher than the inter-PUF distance, FAR increases while FRR reduces, which means better robustness. On the other hand, if the decision is made using a threshold which is significantly lower than the inter-PUF distance, FAR reduces while FRR increases, leading to higher security. By varying the detection threshold, different levels of FAR and FRR can be achieved.

With a small number of transmitters ($\approx 50$), the FAR and FRR for RF-PUF along the equal error line is very close to the pairwise-compared (P.C.) Ring Oscillator PUF [45], and is much better than the other PUFs in Fig. 12. As the number of transmitters increase, the overall error rate increases as the inter-PUF distances reduce. However, depending on the application scenario, the detection threshold within the machine learning framework can be altered, leading to either highly secured or highly robust systems.

## VI. CONCLUSIONS AND FUTURE WORK

In this paper, a conceptual development of RF-PUF is presented along with a feasibility study showing that the inherent RF properties arising from the manufacturing process in a wireless node can be exploited as a strong PUF for device authentication in asymmetric IoT networks without any additional hardware at the transmitters. Using an in-situ machine learning based framework, up to 10,000 transmitters can be detected with about 99% accuracy. The proposed method also eliminates the need for preamble-based or key-based identification of modern IoT nodes and enables low-cost secure authentication using the intrinsic properties embedded in the RF signal that does not have any extra hardware cost at the transmitter. Consequently, the proposed scheme does not consume any additional power at the transmitter side. The receiver, however, requires two neural networks which can be implemented using the on-board microprocessor at a nominal power cost (additional 3-5% overhead when the neural networks are powered on [46]-[47]) which is not significant if the network is asymmetric. As a future direction, more advanced methods of receiver signature compensation will be analyzed along with circuit techniques to implement erasability and certifiability, leading to practical and efficient implementation of the RF-PUF hardware. A formal or experimental validation of the achievable protection degree against different attack models will also be analyzed as a part of the future work. One other research direction involves the stability analysis of RF-PUF in presence of temperature and supply voltage variation.

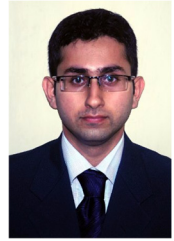
**Baibhab Chatterjee** (S'17) received his B.Tech. degree in Electronics and Communication Engineering from National Institute of Technology Durgapur, India in 2011 and M. Tech. degree in Electrical Engineering from Indian Institute of Technology Bombay, India in 2015. He is currently working towards a Ph.D. degree in the School of Electrical Engineering, Purdue University, West Lafayette, IN, USA. His Industry Experience includes two years as a Digital Design Engineer /Senior Digital Design Engineer in Intel, India and one year as a R&D Engineer in Tejas Networks, India. He was an awardee of the University Gold Medal both during B.Tech. and M.Tech, and is a recipient of the Andrews Fellowship at Purdue University during 2017-18. His research interests include low-power analog, RF and mixed-signal circuit design for secure biomedical applications.

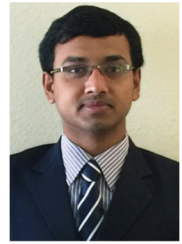
**Debayan Das** (S'17) received the B.E. degree in electronics and telecommunication engineering from Jadavpur University, India, in 2015. He is currently pursuing the Ph.D. degree with the SPARC Lab, Purdue University, West Lafayette, IN, USA. He was an Analog Design Engineer with xSi Semiconductors (start-up) for a year. His research interests include hardware security and mixed-signal IC design. He was a recipient of the IEEE HOST Best Student Paper Award in 2017.

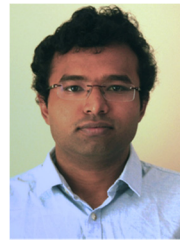
**Shovan Maity** (S'18) received the B.E. degree in electronics and telecommunication engineering from Jadavpur University, India, in 2012, and the M.Tech. degree in electrical engineering from IIT Bombay, in 2014. He was an Analog Design Engineer at Intel, Bangalore, India, from 2014 to 2016. He is currently pursuing the Ph.D. degree in electrical engineering with Purdue University, West Lafayette, IN, USA. His research interests include design of circuits and systems for human body communication, hardware security, and mixed signal circuits design.

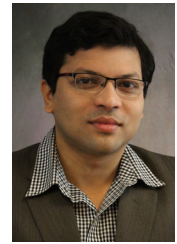
**Shreyas Sen** (S'06-M'11-SM'17) received his Ph.D. degree in Electrical and Computer Engineering from Georgia Tech, Atlanta, USA, in 2011. He is currently an Assistant Professor in School of Electrical and Computer Engineering, Purdue University. He has over 5 years of industry research experience at Intel Labs, Qualcomm and Rambus. His research interests include mixed-signal circuits/systems for Internet of Things (IoT), Biomedical and Security. He has authored/co-authored 2 book chapters, over 100 conference and journal papers and has 13 patents granted/ pending.

In 2018, Dr. Sen was chosen by MIT Technology Review as one of the top 10 Indian Inventors Worldwide under 35 (MIT TR35 India Award), for the invention of using the Human Body as a Wire, which has the potential to transform healthcare, neuroscience and human-computer interaction. Dr. Sen is a recipient of the AFOSR Young Investigator Award 2017, NSF CISE CRII Award 2017, Google Faculty Research Award 2017, Intel Labs Divisional Recognition Award 2014 for industry-wide impact on USB-C type, Intel PhD Fellowship 2010, IEEE Microwave Fellowship 2008, GSRC Margarida Jacome Best Research Award 2007, Best Paper Awards at HOST 2017 and 2018, ICCAD Best-in-Track Award 2014, VTS Honorable Mention Award 2014, RWS Best Paper Award 2008, Intel Labs Quality Award 2012, SRC Inventor Recognition Award 2008 and Young Engineering Fellowship 2005. He serves/has served as an Associate Editor for IEEE Design & Test, Executive Committee member of IEEE Central Indiana Section, ETS and Technical Program Committee member of DAC, CICC, DATE, ISLPED, ICCAD, ITC, VLSI Design, IMSTW and VDAT. Dr. Sen is a Senior Member of IEEE.